\documentclass[
aps,
prb,
reprint,
superscriptaddress,
amsmath,amssymb,
longbibliography,
]{revtex4-1}

\usepackage[version=3]{mhchem} 
\usepackage{graphicx}
\usepackage{dcolumn}
\usepackage{bm}
\usepackage{natbib,hyperref}
\hypersetup{colorlinks=true,allcolors=blue}

\begin{document}

\title{Optical switching of defect charge states in 4\emph{H}-SiC}

\author{D. Andrew Golter}
\author{Chih Wei Lai}
\email[]{chihwei.lai.civ@mail.mil}
\affiliation{U.S. Army Research Laboratory, 2800 Powder Mill Rd, Adelphi, MD 20783}

\date{\today}

\begin{abstract}
We demonstrate optically induced switching between bright and dark charged divacancy defects in 4\emph{H}-\ce{SiC}. Photoluminescence excitation and time-resolved photoluminescence measurements reveal the excitation conditions for such charge conversion. For an energy below ~1.3 eV (above ~950 nm), the PL is suppressed by more than two orders of magnitude. The PL is recovered in the presence of a higher energy repump laser with a time-averaged intensity less than 0.1\% that of the excitation field. Under a repump of 2.33 eV (532 nm), the PL increases rapidly, with a time constant ~30 $\mu$s. By contrast, when the repump is switched off, the PL decreases first within ~100-200 $\mu$s, followed by a much slower decay of a few seconds. We attribute these effects to the conversion between two different charge states. Under an excitation at energy levels below 1.3 eV, \ce{V_{Si}V_C^0} are converted into a dark charge state. A repump laser with an energy above 1.3 eV can excite this charged state and recover the bright neutral state. This optically induced charge switching can lead to charge-state fluctuations but can be exploited for long-term data storage or nuclear-spin-based quantum memory.
\end{abstract}

\maketitle

Optically active point defects (color centers) in wide-band-gap semiconductors can possess long electron spin coherence times ($>$1 ms) and have been considered for use in solid-state quantum sensing and information processing. A prominent example is the negatively charged nitrogen vacancy (\ce{NV^-}) in diamond, in which the electron spin state can be initialized with non-resonant optical excitation and detected via photoluminescence contrast at ambient conditions. Recently, similar defects have been identified in silicon carbide (SiC) for use in wafer-scale quantum technologies \cite{koehl2015}. SiC crystals form in three main polytypes -- 4\emph{H}, 6\emph{H} (hexagonal), and 3\emph{C} (cubic) -- offering a broad range of defects \cite{iwamoto2015} that can act as potential spin qubits, among which are the carbon antisite-vacancy pair (\ce{C_{Si}V_C}) \cite{umeda2006,steeds2009,castelletto2013,szasz2015}, the nitrogen vacancy (\ce{N_CV_{Si}}) \cite{vonbardeleben2015,vonbardeleben2016,zargaleh2016}, the silicon monovacancy \ce{V_{Si}^-} \cite{baranov2011,widmann2014,simin2015,simin2016}, and the neutral divacancy (\ce{V_{Si}V_{C}^0}) \cite{son2006,koehl2011,christle2015}.


\begin{figure}[htbp]
\centering
\includegraphics[width=\linewidth]{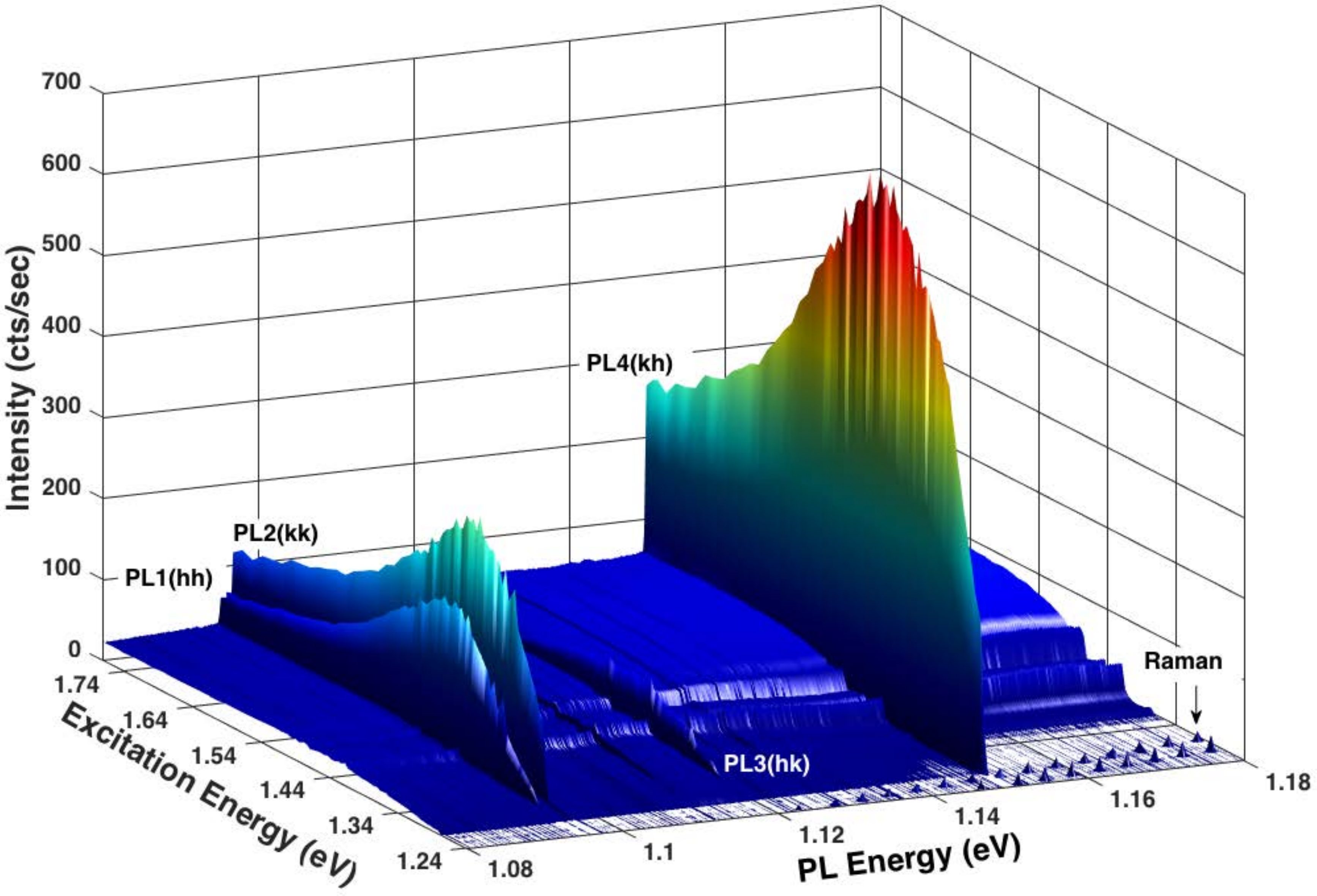}
\caption{PLE spectra under an excitation energy ($E_{ex}$) from 1.24 to 1.77 eV at T = 10 K. The four major zero-phonon-line (ZPL) emissions are labeled as PL1($hh$), PL2($kk$), PL3($hk$), and PL4($kh$). The broad background emissions are from the phonon side bands of \ce{V_{Si}V_C} and \ce{V_{Si}}. The arrow indicates one of the Raman LO and TO peaks.}
\label{fig:ple2d}
\end{figure}

In this work, we focus on the \ce{4\emph{H}-SiC} polytype, in which the two adjacent carbon and silicon vacancies of \ce{V_{Si}V_C^0} organize in either axial ($hh$, $kk$) or basal ($hk$,$kh$) configurations as a result of varying lattice sites and orientations. We identify the optimal pump laser energy by using photoluminescence excitation (PLE) measurements (see Methods). A unique zero phonon line (ZPL) is associated with each configuration and is labeled as PL1--PL4 with emission energies (wavelengths) $E_{ZPL}\approx$1.096--1.150 eV ($\lambda_{ZPL}\approx$1078--1131 nm) (Fig. \ref{fig:ple2d}). The ZPL intensities ($I_{ZPL}$) are proportional to the phonon-side-band (PSB) absorption and peak under an excitation energy $E_{ex}\sim$1.4 eV. The PSB absorption is generally finite over hundreds of meVs and only becomes negligible for near-resonant excitations. Therefore, when $E_{ex}$ is tuned toward $E_{ZPL}$, $I_{ZPL}$ is expected to decrease gradually with slowly decreasing PSB absorption. Surprisingly, $I_{ZPL}$ decreases precipitously by more than two orders of magnitude for $E_{ex}\lesssim$1.278 eV (970 nm). Moreover, the threshold-like drop of $I_{ZPL}$ occurs at distinct $E_{ex}$ for defects at inequivalent lattice sites.

\begin{figure}[htbp]
\centering
\includegraphics[width=\linewidth]{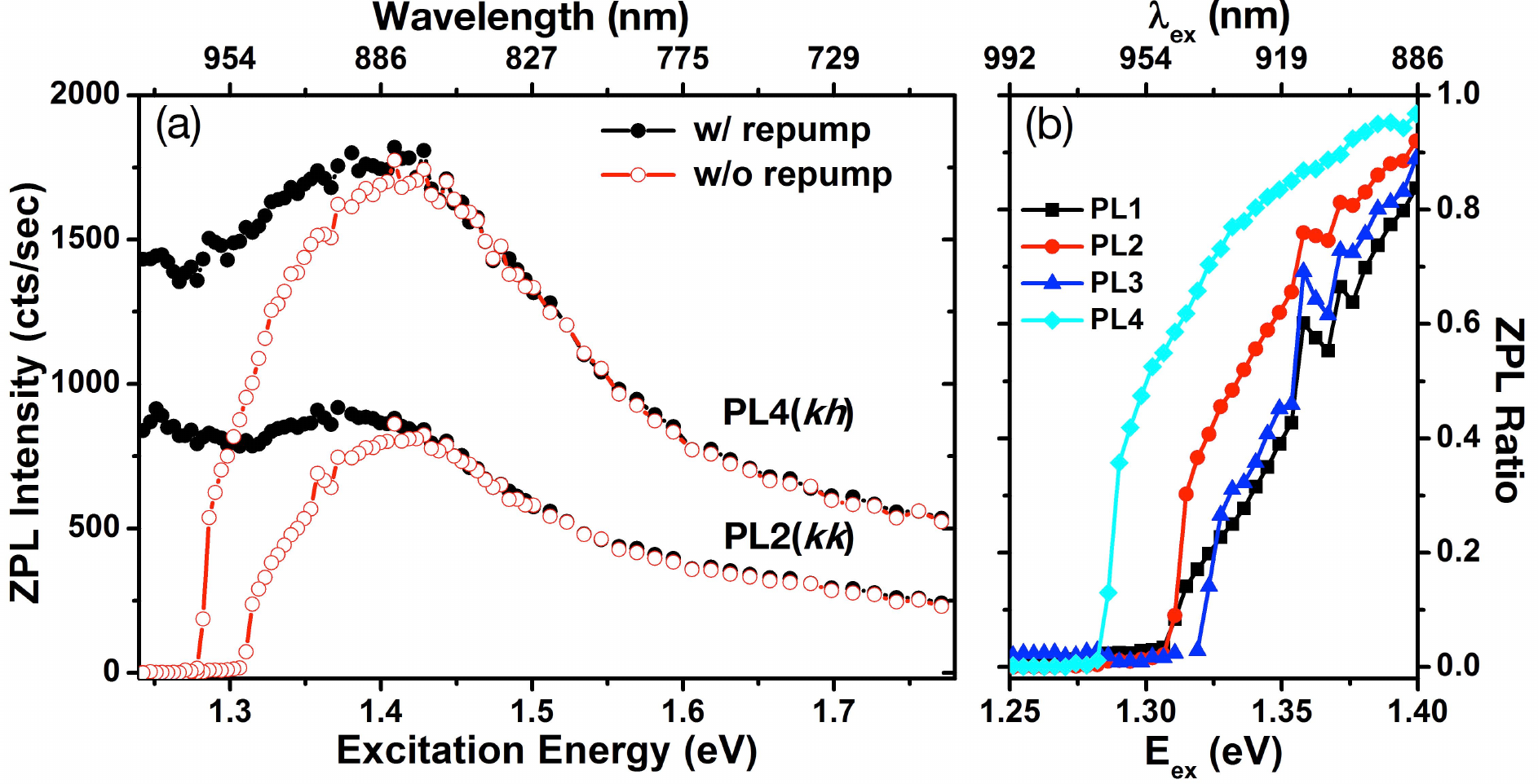}
\caption{(a) Spectrally integrated ZPL intensities ($I_{ZPL}$) of PL2($kk$) axial and PL4($kh$) basal defects with (black) and without (red) the 532 nm repump laser as a function of the pump (excitation) energy. Pump laser power is maintained at 20 mW, while the repumping 532-nm laser power is 0.2 mW. (b) Ratios of $I_{ZPL}$ with and without the repump (ZPL ratio) for four types of divacancies as labeled in Fig. \ref{fig:ple2d}.}
\label{fig:ple}
\end{figure}

To better determine the energy at which the ZPL is suppressed precipitously, we plot the spectrally-integrated $I_{ZPL}$ for individual ZPL in Fig. \ref{fig:ple}. We also find that that this PL suppression at lower energies is reversed by the addition of a 2.33 eV (532 nm) repump field with a time-averaged intensity about 0.1\% that of the excitation field, as shown by the black curves for PL2 and PL4 in Fig. \ref{fig:ple}a. The four ZPLs vary in $I_{ZPL}$, likely due to a combination of difference in defect density, radiative recombination efficiency, and polarization- or dipole-orientation-dependent optical collection efficiency. Thus, we plot the ratio of $I_{ZPL}$ with and without repump to better display the transitions for all four defect types PL1--PL4 on the same plot (Fig. \ref{fig:ple}b). The excitation energy at which ZPL is dramatically suppressed ranges from approximately 1.28 to 1.32 eV.

\begin{figure}[htbp]
\centering
\includegraphics[width=\linewidth]{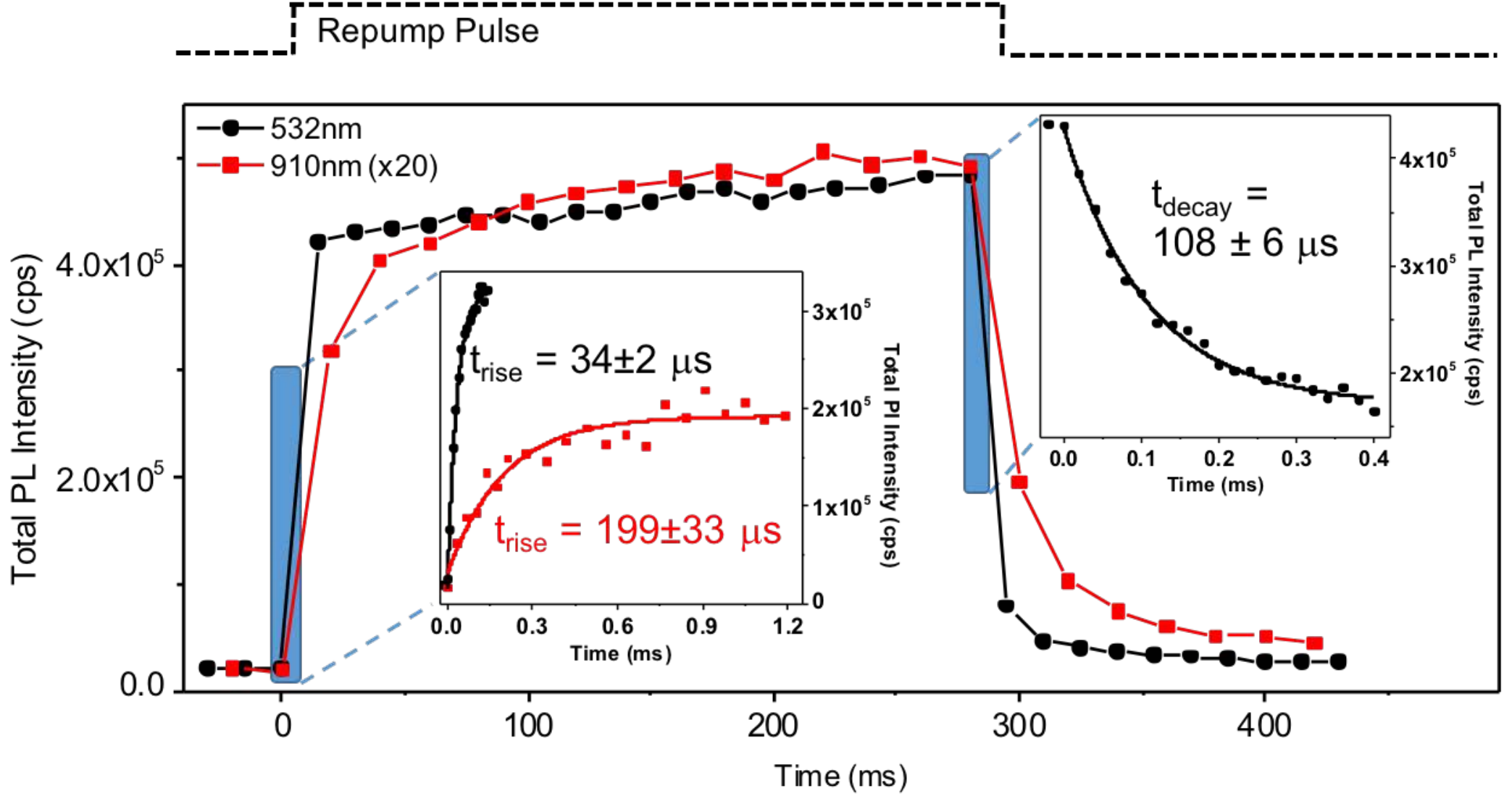}
\caption{Time-resolved measurement of the total PL from \ce{V_{Si}V_C^0} during the repump and suppression processes. The pump laser is on continuously while the repump laser is pulsed. Insets show the fast initial growth in PL when the repump laser is turned on and the suppression when the repump is turned off. Results are shown for a repump at 532 nm (black) as well as at 910 nm (red). The 910 nm curve is scaled (x20) and offset. The rise time of the repump process is shorter for the higher energy repump. After an initial rise over 10-100 $\mu$s, PL reaches a plateau with a much slower growth (100's of ms), likely due to defects near the periphery of the excitation spot which see a significantly lower excitation intensity. The decay after the 910 nm repump is identical to that for 532 nm.}
\label{fig:dynamics}
\end{figure}

Next, we determine the dynamics of the repump and suppression processes with time-resolved PL measurements (Fig. \ref{fig:dynamics}). The sample is excited continuously by a laser at 1.27 eV (976 nm). To switch the defects between the bright and dark states, we use a pulsed repump laser with a time-averaged intensity $<$0.1\% that of the excitation field. Under a repump of 2.33 eV (532 nm), the PL increases rapidly, with a time constant $\approx$30 $\mu$s. In contrast, when the repump is switched off, the PL decreases first within $\approx$100--200 $\mu$s, followed by a much slower decay of a few seconds. The non-exponential rise and fall of the PL become more evident under a repump of 1.362 eV (910 nm). In this case, a rapid sub-ms surge in PL is followed by a gradual increase over a few ms, while the PL decay remains similar to that observed under the 532-nm excitation. 

The fast initial rise times depend on the repump power. The repump power that was chosen for the plots in Fig. 3 was close to saturation for this process. For instance, we found that decreasing the 532nm repump power by almost an order of magnitude increased the rise time by about a factor of two. From our model we would expect the repump process to depend on both repump energy and intensity. However accurate determination of such energy and power dependence is not feasible from our ensemble measurements, owing to the fact that the local intensity experienced by each defect varies for different regions of the excitation spot.


We attribute these effects to the conversion between two different charge states, as observed in \ce{NV^0}/\ce{NV^{-}} centers in diamond \cite{beha2012,siyushev2013}. Under an excitation at energy levels below 1.3 eV, \ce{V_{Si}V_C^0} are converted into a dark charge state, where the system becomes trapped. A repump laser with an energy above 1.3 eV can excite this charged state and recover the bright neutral state. This conversion is most effective with a repump above 1.4 eV.

\begin{figure}[htbp]
\centering
\fbox{\includegraphics[width=0.95\linewidth]{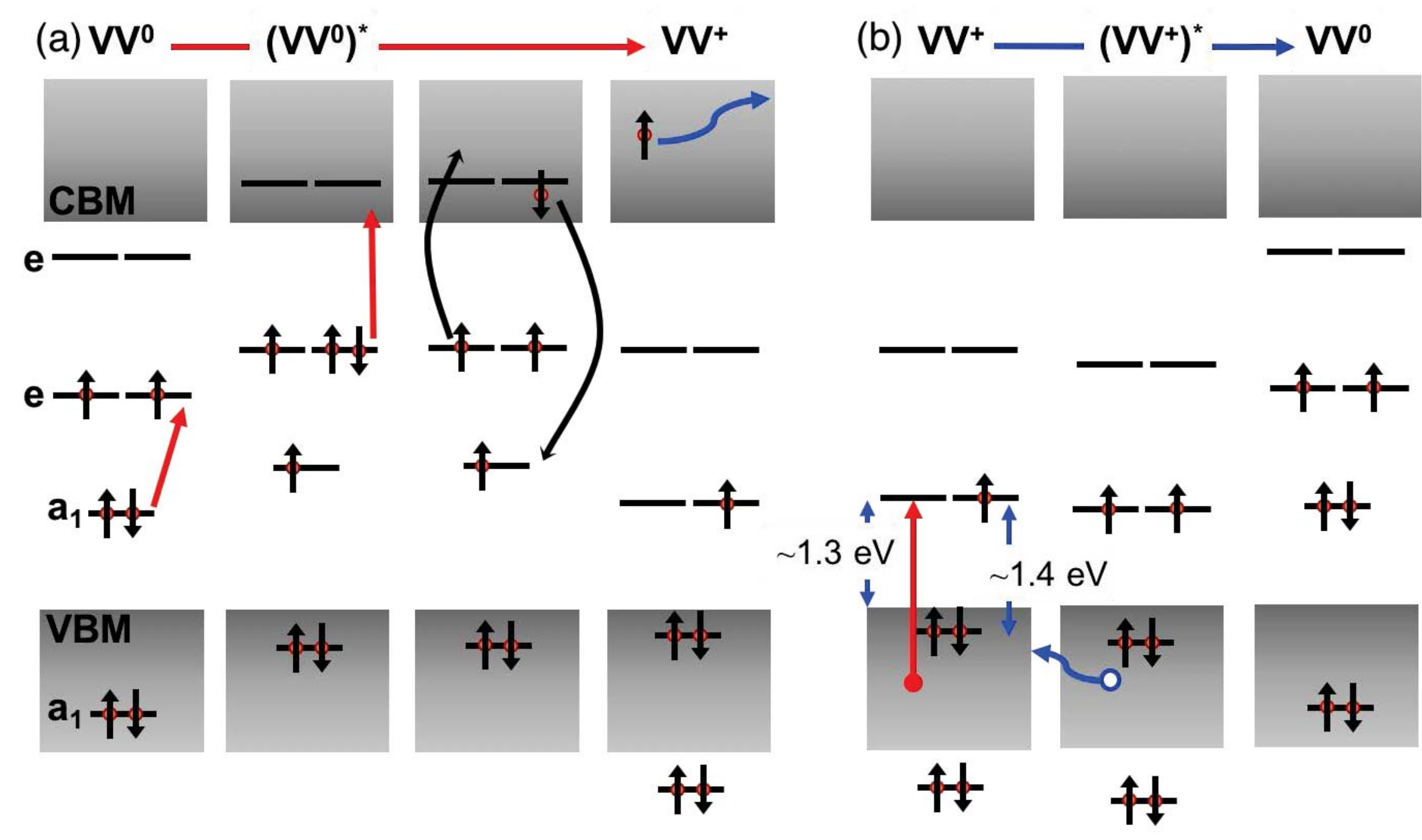}}
\caption{Schematics of the charge conversion process for the neutral and positive states of the \ce{V_{Si}V_C} defect. (a) \ce{VV^0} to \ce{VV^+} conversion involves two photons and an Auger process that release sufficient energy to ionize an electron from the defect. (b) The $a_1$ orbital of \ce{VV^+} lies in the valence band. When an electron is excited from the continuum states in the valence band (or $a_1$ orbital, not shown) to the $e$ orbital, the hole migrates away from the defect, converting the \ce{V_{Si}V_C} center back to the neutral state.}
\label{fig:model}
\end{figure}

We consider the conversion between the neutral and positive charge (\ce{V_{Si}V_C^+}) states (Fig. \ref{fig:model}). In the \ce{V_{Si}V_C} divacancy, there are two $a_1$ and two $e$ states, which are formed from by the six dangling bonds. The position of these defect energy levels varies with the occupation of the states, as wells as with the relative crystalline position of the divacancy pair. In the neutral charge state, four electrons occupy the two $a_1$ states, and two electrons occupy the lower, degenerate $e$ state \cite{gordon2015,iwata2016}. The upper $a_1$ and two $e$ levels are within the gap (Fig. \ref{fig:model}a), resulting in the neutral divancancies undergoing atomic-like transitions. A charge neutral defect can be ionized and converted into a positive charge state through two-photon absorption, followed by the Auger process (Fig. \ref{fig:model}a). In \ce{V_{Si}V_C}, the defect energy levels of the non-zero charge states remain largely unknown, though the formation energies of stable charged states, namely, $+$, $0$, $-$, and $-2$, have been calculated \cite{iwata2016}. For the positive charge state, we envision that the upper $a_1$ state lies about 0.1--0.15 eV below the valence band maximum (VBM), while the lower $e$ state is likely about 1.3 eV above the VBM (Fig. \ref{fig:model}b). 

Under an excitation energy below 1.3 eV but above the \ce{VV^0} ZPL transition, the conversion from \ce{VV^0} to \ce{VV^+} remains effective; however, the defect becomes 'trapped' in the \ce{VV^+} because the optical excitation of an electron out of the valence band is suppressed. Under an excitation of energy above 1.3 eV, an optically excited electron occupies one of the lower $e$ states, while the hole rapidly migrates away from the defect (i.e., an electron is captured) (Fig. \ref{fig:model}b). As a result, the defect reverts to the neutral charge state. This charge reversion involves only one-photon absorption and is expected to occur at a higher rate than ionization, which is consistent with our measurements (insets in Fig. \ref{fig:dynamics}). When this hole migration rate is much higher than that of the radiative recombination, the positive charge state is optically dark, as is observed experimentally. The aforementioned electron-capture rate increases with increasing optical excitation energy from approximately 1.28 to 1.4 eV, and becomes nearly constant for excitation energies exceeding the $a_1$-$e$ transition energy, about 1.4 eV. The hypothesis of an increasing electron-capture rate with increasing repump energies is supported by the distinct switching dynamics under a 532-nm or 910-nm repump (Fig. \ref{fig:dynamics}).

In this model, we expect the energy difference between the upper $a_1$ orbital and VBM to vary for defects in inequivalent lattice sites, resulting in distinct 'threshold' excitation energies as shown in Fig. \ref{fig:ple}. By contrast, the energy gap between the upper and lower $a_1$ orbitals should be insignificant as suggested by the experimental observation that PL1--PL4 all peak around $E_{ex}\approx$ 1.4 eV in the absence of repump. Further calculations of the defect energy levels based on density function theory (DFT) should verify the model proposed here.
 
While preparing this manuscript, we learned that Wolfowicz et al. have also observed optically induced switching between bright and dark states \cite{wolfowicz2017}. They model this conversion based on switching between neutral and negative charge (\ce{V_{Si}V_C^-}) states. The cycling between neutral and negative charge states requires the inclusion of other shallow donors surrounding the divancancies. Our experimental results do not preclude such a scenario. To determine whether the dark state is positively or negatively charged, it is necessary to examine optically induced switching of charge states of single defects and compare with accurate DFT calculations of the electronic structures of these charge states.  


In conclusion, we demonstrate that \ce{V_{Si}V_C} divacancies can become charged via optical excitation. In diamonds, both \ce{NV^0} and \ce{NV^-} are optically active with identifiable ZPLs. By contrast, the PLE and PL measurements in \ce{V_{Si}V_{C}} in 4\emph{H}-\ce{SiC} suggest that the charged \ce{V_{Si}V_{C}^-} or \ce{V_{Si}V_{C}^+} states are optically dark. This optically induced charge switching can lead to charge-state fluctuations but can also be exploited for long-term data storage or nuclear-spin-based quantum memory, as shown for NV centers in diamonds \cite{dhomkar2016}.

\section*{Methods}
\paragraph*{Sample.}
The sample is a high-purity semi-insulating (HPSI) 4\emph{H}-silicon carbide substrate purchased from Norstel. The \ce{V_{Si}V_C} divacancies are naturally formed without additional electron/proton irradiation or annealing. The density of \ce{V_{Si}V_C} defects is estimated to be about $10^{14}$ to $10^{16}$ cm$^{-3}$. Similar optical induced charge switching effects are also observed in HPSI SiC substrate purchased from Cree, Inc. 
 
\paragraph*{Setup.}
Excitation and repump lasers are focused to a $\sim\mu$m$^2$ area on the sample via a microscope objective with NA = 0.75. A tunable single-frequency laser (M Squared Lasers SolsTiS) is used for the excitation with a constant time-averaged power of 20 mW ($\pm$0.2 mW) for the PLE measurement. The same objective collects the ensemble PL, which is filtered from the reflected excitation and repump lasers via a longpass filter. PL spectra are measured with a spectrometer equipped with a liquid-nitrogen-cooled InGaAs array. Time-resolved PL is measured by coupling spectrally-integrated ZPL and phonon-side-band emissions from \ce{V_{Si}V_C^0} divacancies to a superconducting nanowire single photon detector (Single Quantum Eos). The repump laser pulses are created using an acousto-optic modulator.

\section*{Acknowledgements}
This work was supported by Office of Secretary of Defense, Quantum Science and Engineering Program. The authors thank D. E. Taylor, A. Beste, B. VanMill, and D. D. Awschalom for discussions and providing SiC samples.


\end{document}